\documentclass{article}

\usepackage{PRIMEarxiv}

\usepackage[utf8]{inputenc} 
\usepackage[T1]{fontenc}    
\usepackage{hyperref}       
\usepackage{url}            
\usepackage{booktabs}       
\usepackage{amsfonts}       
\usepackage{nicefrac}       
\usepackage{microtype}      
\usepackage{lipsum}
\usepackage{fancyhdr}       
\usepackage{graphicx}       
\usepackage{amsmath}
\graphicspath{{media/}}     

\title{Research on Multi-Agent Communication and Collaborative Decision-Making Based on Deep Reinforcement Learning
}

\author{
  Zeng Da \\
  University of Electronic Science and Technology of China \\
  \texttt{809235195@qq.com} 
}

\begin{document}
\maketitle

\begin{abstract}

In a multi-agent environment, the continuous change of each agent's strategy will lead to the non-stationarity of the multi-agent environment, so the reinforcement learning problem in a multi-agent environment is usually modeled as a decentralized partially observable Markov decision Process (Decentralized Partially Observable Markov Decision Process, Dec-POMDP), which brings challenges to the cooperation between agents. In order to overcome and alleviate the non-stationarity of the multi-agent environment, the mainstream method is to adopt the framework of Centralized Training Decentralized Execution (CTDE). This thesis is based on the framework of CTDE, and studies the cooperative decision-making of multi-agent based on the Multi-Agent Proximal Policy Optimization (MAPPO) algorithm for multi-agent proximal policy optimization.

(1) In order to alleviate the non-stationarity of the multi-agent environment, a multi-agent communication mechanism based on weight scheduling and attention module is introduced. Different agents can alleviate the non-stationarity caused by local observations through information exchange between agents, assisting in the collaborative decision-making of agents. The specific method is to introduce a communication module in the policy network part. The communication module is composed of a weight generator, a weight scheduler, a message encoder, a message pool and an attention module. Among them, the weight generator and weight scheduler will generate weights as the selection basis for communication, the message encoder is used to compress and encode communication information, the message pool is used to store communication messages, and the attention module realizes the interactive processing of the agent's own information and communication information .

(2) In the CTDE framework, global information is introduced during centralized training to alleviate environmental non-stationarity. In the MAPPO algorithm, the input of the centralized value network contains global information, and the processing of global information has an impact on the estimation of the value function. This thesis proposes a global information processing module based on the attention mechanism and deep and shallow feature processing. The global information and the local observation information of each agent are input into the attention module to obtain the simplified feature information, after deep and shallow layer feature processing, it is used as the input of the value network.

Combining the above improvements, this thesis proposes a Multi-Agent Communication and Global Information Optimization Proximal Policy Optimization(MCGOPPO)algorithm, and conducted experiments in the StarCraft Multi-Agent Challenge (SMAC) and the Multi-Agent Particle Environment (MPE). The experimental results show that the improvement has achieved certain effects, which can better alleviate the non-stationarity of the multi-agent environment, and improve the collaborative decision-making ability among the agents.
\end{abstract}


\section{Introduction}

In today's Internet era, a variety of games have emerged on various platforms, and they are gradually entering Public life, become the seasoning of people's entertainment. Generally speaking, there are two kinds of characters in games, one is the Character controlled by the player, the other is the Non-Player character (NPC) in the game. One problem is that the running logic of the NPC basically follows the logic preset by the game developer when designing the game. The interaction with the player character is very mechanical and fixed, and if the NPCs were more intelligent, it would greatly enhance the player's experience and make the experience of the game world more realistic for the player.

Deep reinforcement learning in the field of artificial intelligence, with its strong autonomous learning ability, is very suitable for swimming the problem of intelligence in drama [8 -- 10]. A typical application scenario is the game Intelligent Bot (AIBot), which can realize the functions of hanging up the phone, acting as a sparring opponent, assisting developers in testing and so on. For example, in competitive games, if a player drops out, the AIbot of the game can be used to replace the player, which makes the experience of the two sides better. It can also be used as a sparring opponent, allowing players to fight against AIbot and compete with human players after skillful operation to help improve players' operation level. In this field of application, Jiwu AI [11] developed by Tencent AILab based on King of Glory has achieved a high-level AIbot. The ultimate AI can reach the level of professional players. In addition, intelligent AIbot can also better simulate the operation and strategy of real players, which is helpful for game developers to conduct simulation tests on character abilities after game development, and will be of great help to game designers in numerical design.

\section{MAPPO}
Multi-agent near-end strategy optimization algorithm, MAPPO for short, adopts a centralized value function approach to consider global information, which belongs to the CTDE framework. It makes each individual PPO agent cooperate with each other through a global value function.Each agent in MAPPO generates an action based on local observation information and strategies to maximize the fold Withholding bonus.
\begin{equation}
J(\theta)=\mathrm{E}_{a^t, s^t}\left[\sum_t \gamma^t R\left(s^t, a^t\right)\right] 
\label{MAPPOreward}
\end{equation} 

MAPPO is composed of an Actor network and Critic network, with a mapping to learn for the value function $V_\phi$ (S→R). The strategy function $\pi_\theta$ learns a mapping from the observed distribution to a range or a mapping to the mean and variance of the action to the Gaussian function for subsequent sampling actions.Actor network optimization objective is:
\begin{equation}
\begin{split}
&L(\theta)=\left[\frac{1}{B n} \sum_{i=1}^B \sum_{k=1}^n \min \left(r_{\theta, i}^{(k)} A_i^{(k)}, \operatorname{clip}\left(r_{\theta, i}^{(k)}, 1-\epsilon, 1+\epsilon\right) A_i^{(k)}\right)\right] \\
&+\sigma \frac{1}{B n} \sum_{i=1}^B \sum_{k=1}^n S\left[\pi_\theta\left(o_i^{(k)}\right)\right], \text { where } r_{\theta, i}^{(k)}=\frac{\pi_\theta\left(a_i^{(k)} \mid o_i^{(k)}\right)}{\pi_{\theta \text {old}}\left(a_i^{(k)} \mid o_i^{(k)}\right)} .
\label{MAPPOActorequa}
\end{split}
\end{equation}

B indicates the size of batch\_size, and n indicates the number of agents. r is the scale coefficient of importance sampling. Represents the advantage function, which is used to measure the rationality of selecting a specific action A under a certain state s. $\epsilon$ represents the clipping coefficient, S represents the entropy of the strategy, and $\sigma$ is a hyperparameter that controls the entropy coefficient.Critic network optimization goal is:

\begin{equation}
\begin{split}
&L(\phi)=\frac{1}{B n} \sum_{i=1}^B \sum_{k=1}^n\mathrm { max } \left\{\left(V_\phi\left(s_i^{(k)}\right)-\hat{R}_i\right)^2\right.. ,\\
&\left.\left[\operatorname{clip}\left(V_\phi\left(s_i^{(k)}\right), V_{\phi_{\text {old }}}\left(s_i^{(k)}\right)-\varepsilon, V_{\phi_{\text {old }}}\left(s_i^{(k)}\right)+\varepsilon\right)-\hat{R}_i\right]^2\right\}
\label{MAPPOCriticequa}
\end{split}
\end{equation}

B indicates the size of batch\_size, and n indicates the number of agents. V is the value function, $\hat{R}_i$ is the discount reward.

\section{MCGOPPO}

The algorithm in this chapter is improved based on IPPO algorithm and MAPPO algorithm. In order to strengthen the collaboration and cooperation among multiple agents, a communication module is introduced between independent agents to assist the collaborative decision-making of agents through information sharing and circulation between agents.

The communication module is divided into two sub-modules, one of which consists of weight generator and weight scheduler
Its function is to improve the efficiency of communication. The weight generator generates the weight of the corresponding agent according to the input information of each agent, and then stores it in the weight scheduler and carries out normalization processing for the selection of communication objects by the agent. The other submodule is composed of the attention module, whose function is to filter the communication information and extract the concise and important communication content. In the multi-agent system environment, due to the partial observability of the environment, a single agent can only obtain partial observation information when performing actions. Therefore, the introduction of inter-agent communication can simultaneously share information and extract key information through attention mechanism to assist target selection, optimize action selection, and improve the level of collaborative decision making among multi-agents.

The overall framework of the near-end strategy optimization algorithm based on multi-agent communication and global information optimization still adopts the actor-critic network framework similar to MAPPO, which is composed of distributed Actor network, centralized Critic network and sample pool.

Distributed Actor network represents each agent and is responsible for the interaction with the environment. The input of Actor network is the local observation information of the agent, and the output is the selective action of the agent. Actor network is composed of weight generator, weight scheduler, message encoder, message pool, attention module and action selector for communication.

More specifically, compared with MAPPO algorithm, the improvement of Actor network lies in the introduction of communication between multiple agents. Through the communication between agents, the information exchange between agents can be improved. On the one hand, the non-stationiness in multi-agent environment can be reduced; on the other hand, more abundant information can be used to assist Actor network. Among them, a single Actor network processes the execution of the corresponding single agent and distinguishes the decision-making of the agent. The communication part is divided into two sub-modules, one is the communication scheduling module based on weight scheduling, and the other is the communication message processing module based on attention mechanism. The communication scheduling module is composed of message encoder, message pool, weight generator and weight scheduler, which is responsible for the compression encoding of communication messages and the generation and allocation of corresponding scheduling weights. The communication message processing module is composed of the attention module, which is mainly responsible for processing the local observation information of the agent together with the communication message, and output the feature information to the subsequent action selector.

The goal of centralized Critic network is to optimize action selection and weight, and assist to update action selection of Actor network. The input is the sample obtained from the sample pool (including the joint set of agent local observation, action selection and reward), as well as global information, and the output is the Value value function.
The centralized Critic network here does not mean that there is only one Critic network, but that the input of Critic network contains global information, which belongs to a form of centralized training under the CTDE framework. Different from MAPPO algorithm, when dealing with global information, this paper introduces the attention unit to deal with it. Later, the processed global information will continue to undergo deep and shallow feature processing, and then input it into the Critic network to calculate the value function and assist the update of the Actor network.

The overall training process is to interact with Actor network and environment to obtain local observation information, local observation
The information o is compressed and coded by the message encoder to get the communication message m and write it into the message pool. At the same time, the local observation information o is also generated by the weight generator to generate the weight w and then input it to the weight scheduler. When two Actor networks communicate with each other, one of them will select the communication object according to the weight scheduler. In this way, the communication information m in the message pool of other agents is read, and then the local observation information o and communication information m are input into the attention module. The information filtered by the attention module is the characteristic information c obtained by the current agent after integrating the communication information. The characteristic information c is input into the action selector and the action selection a of the agent is output. Actor network then interacts with the environment through the output action and gets the corresponding reward. After multiple cycles, the collected observation information o, action information a, and reward r compose samples and input them into the sample pool, and then add the global information s. After passing through the attention unit and the Critic feature processing layer, input them into the critic network, and calculate the output Value function.

\subsection{Multi-agent communication module based on weight scheduler and attention mechanism}

This section introduces the multi-agent communication module based on weight scheduling and attention mechanism, which is divided into two modules Part.The first is the communication scheduling module based on weight scheduling, and the second is the message processing module based on the attention mechanism. For example, when every agent needs to communicate with each other, the communication between two agents will take up a large amount of communication bandwidth, which will appear very redundant in a limited bandwidth environment. Moreover, excessive redundant information is easy to introduce noise information, which will indirectly affect the subsequent decision-making of the agent. Therefore, a communication scheduling module based on weight scheduling is introduced to solve this problem.

At the same time, the follow-up processing of communication information is not simply related to the local view of the agent itself The measurement information is superimposed. In the absence of communication, the decision-making basis of each agent is its own local observation information. After the addition of communication, the introduced communication information is essentially the characteristics of the local observation information of other agents after compression and coding processing. The content form of these two parts and the meaning of the information represented by them overlap to a certain extent. In order to avoid the redundancy between the introduced communication information and its own observation information, this paper proposes a communication message processing module based on attention mechanism.

Message encoder: consists of two MLP layers. The input is the local observation information oi of the agent, and the output is The communication information mi after encoding and compression will be written into the message pool for storage; The process can be abstracted as a mapping.

Weight generator: It consists of three MLP layers. The input is the local observation information oi of the agent, and the output is the weight $w_i$. Its essence is a value, which determines the probability of the message being selected in the subsequent communication. The process can be abstracted as a mapping.

Weight scheduler: Essentially, it is a SoftMax layer.  The input is the set of weight information of each agent $W=[w_1,\dots,w_n]$, and the output is the final scheduling weight set $W'=[{w_1}' , \dots , {w_n}']$.  This module can be viewed as a mapping from the weight w (generated by $f^i_{wg}$) to w0 of all agents.  The specific processing method is to sort directly according to the normalized size of the generated weight w, and select the message corresponding to the weight at the top of the ranking as the communication message to be delivered.  The process can be abstracted as a mapping.

Communication, according to the weight scheduler processing, select the message pool stored in the communication message $m_j$ (here The subscript j of represents that the communication message is obtained from the message pool in agent j after scheduling), which is the final communication message $c_i$ obtained by the current agent i through communication. The message is then forwarded to a subsequent attention module for further processing.

\begin{equation}
\begin{aligned}
Attention(Q_i,K_j,V_i)= SoftMax\left(\frac{Q_i{K_j}^T}{\sqrt[]{d_k}}\right)V_i
\end{aligned}
\label{attij}
\end{equation}

Where, the subscript of $Q_i$,$K_j$ and $V_i$ represents the conversion source of the matrix. Subscript is i, indicating that the matrix is converted from oi, the local observation information of agent i; subscript is j, indicating that the matrix is converted from $m_j$, the communication information in the message pool of agent j. The operational meaning of this attention mechanism corresponding to the actual multi-agent environment is that, by extracting the communication information $m_j$ (namely $c_i$) of agent j and the more closely related information oi of agent i's own local observation information, as the useful part obtained by agent i in this communication process, the output is $z_i$. Then enter it into the action selector to continue processing.

\subsection{Global information optimization processing module based on attention mechanism and deep and shallow feature processing}

The reason for deep and shallow feature processing in this paper is that: because the whole feature information contains the Agent's own information, local information of friends and enemies, mobile information and Agent\_id information, the importance of which information is not uniform. Therefore, this paper proposes the method of deep and shallow feature processing, hoping to do a differentiated processing of different parts of information. Enemy information for the current agent, is closely related to target selection, so it is relatively more important to do deep processing, friends and their own information to do shallow processing. 

The input of deep and shallow layer feature processing is the feature information processed by the attention unit before, which is divided into two parts at first, and one part is enemy information, which will be input into the three FC layers for further processing. The other part is the Agent's own information, local information of friends and enemies, mobile related information and Agent\_id related information, only input to a FC layer for shallow processing. Finally, the features of the two parts were spliced together to get the final feature information and input it into the Critic network.

\section{Experiments}

SMAC Environment is an open source multi-agent reinforcement learning field based on the StarCraft II game. It has the characteristics of huge observation space and action space, local observation and long-term decision making. Especially with so many game units to manipulate, it is challenging to study and try to come up with models with better synergies and easier convergence. It also offers an interesting set of micro challenge scenario maps, which contain more than two dozen challenge scenarios inspired by the StarCraft Master Challenge mission released by Storm Snow, and are designed to assess how independent agents learn high-level collaboration and micromanagement techniques. Specifically, there are two camps in the SMAC scenario, one controlled by the researchers themselves and the other controlled by a carefully designed non-learning heuristic algorithm, and victory is conditional on defeating and destroying all units of the opposing camp. Based on the level of built-in AI and the difference between the two camps: easy, hard and super hard.

In order to study the effectiveness of the algorithm proposed in this paper, six maps of 2s3z, MMM, bane\_vs\_bane, 2s\_VS\_1sc, 5m\_VS\_6m and corridor are selected to verify the algorithm. The information of the agent, camp type and difficulty degree corresponding to each map is shown in Table 4-2: A total of 6 maps were used for the experiment, and two types of symmetric map and asymmetric map were selected according to camp type. For each type of map, three difficulty levels of simple, difficult and super difficult maps were selected respectively according to the difficulty level. The evaluation results of the experiment are composed of the win rate curve and the reward curve. The advantages and failures of the algorithm in different maps are judged according to the convergence curve of the win rate and the reward growth curve. Four algorithms are adopted In experiments, IPPO algorithm and MAPPO algorithm are the baseline algorithm in comparison, and QMIX algorithm is the algorithm most commonly used for comparison in mainstream research on multi-agent reinforcement learning. MCGOPPO algorithm is the final improved algorithm which combines the improvement of multi-agent communication based on weight scheduling and attention module and the improvement of global information optimization based on attention mechanism and deep and shallow feature processing. In the experimental graph, IPPO algorithm is represented by a green dotted line, MAPPO algorithm by a blue dotted line, QMIX algorithm by a purple dotted line, and MCGOPPO algorithm by a red solid line.

First, we conducted the scene experiment of symmetrical camp maps in SMAC environment, namely 2s3z, MMM and bane\_vs\_bane maps, in which the difficulty of the maps gradually increased.
\begin{figure}[h]
\includegraphics[width=10cm]{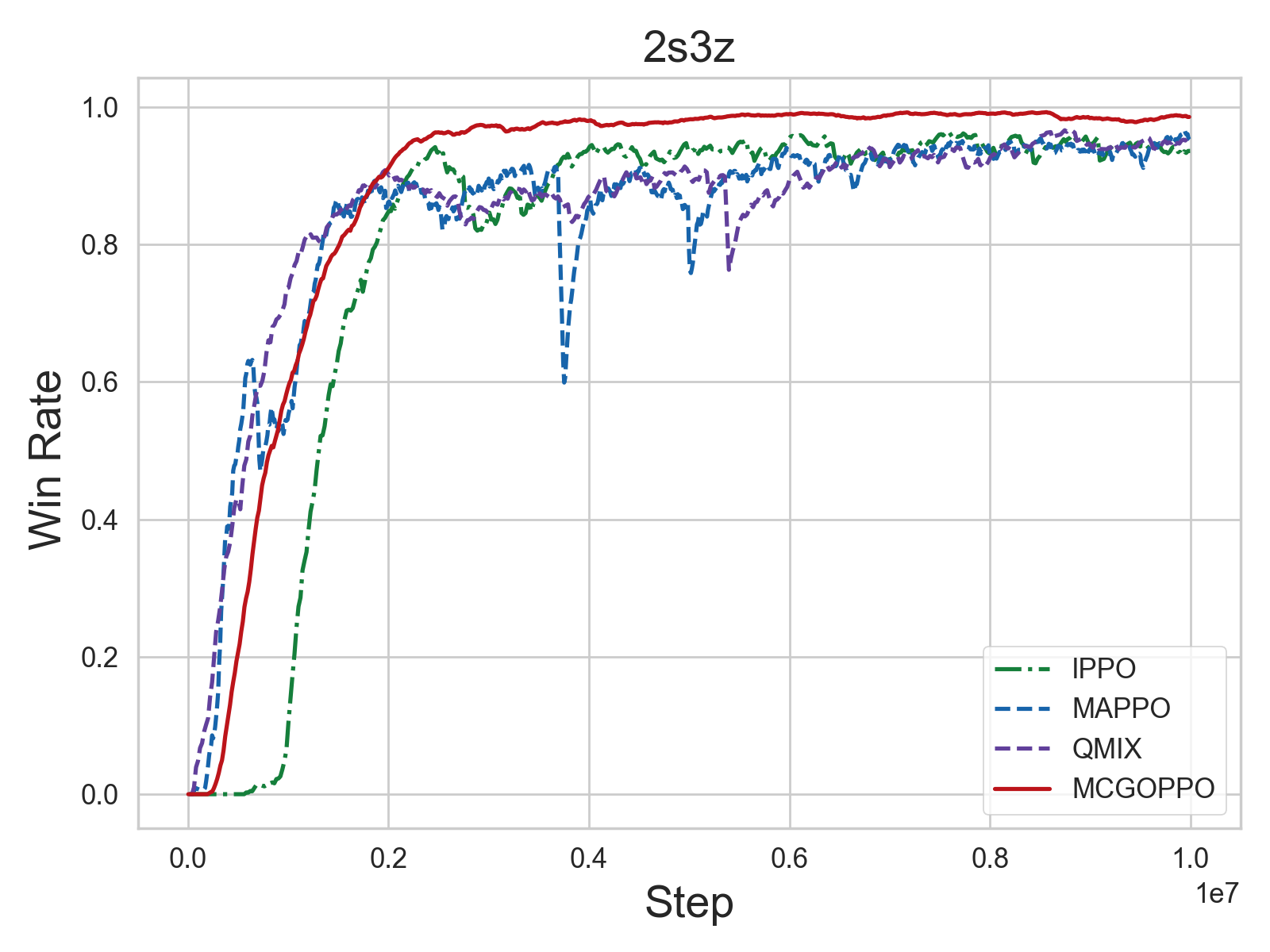}
\caption{Experimental win ratio of scene 2s3z in SMAC}
\label{2s3z_winrate}
\end{figure}

MAPPO algorithm and QMIX algorithm in the early win rate increase the fastest, but in a million It was overtaken by MCGOPPO algorithm around the step, and then MCGOPPO algorithm kept the lead. It can be considered that MCGOPPO algorithm is superior to other algorithms on this map, with a good improvement.

\begin{figure}[h]
\includegraphics[width=10cm]{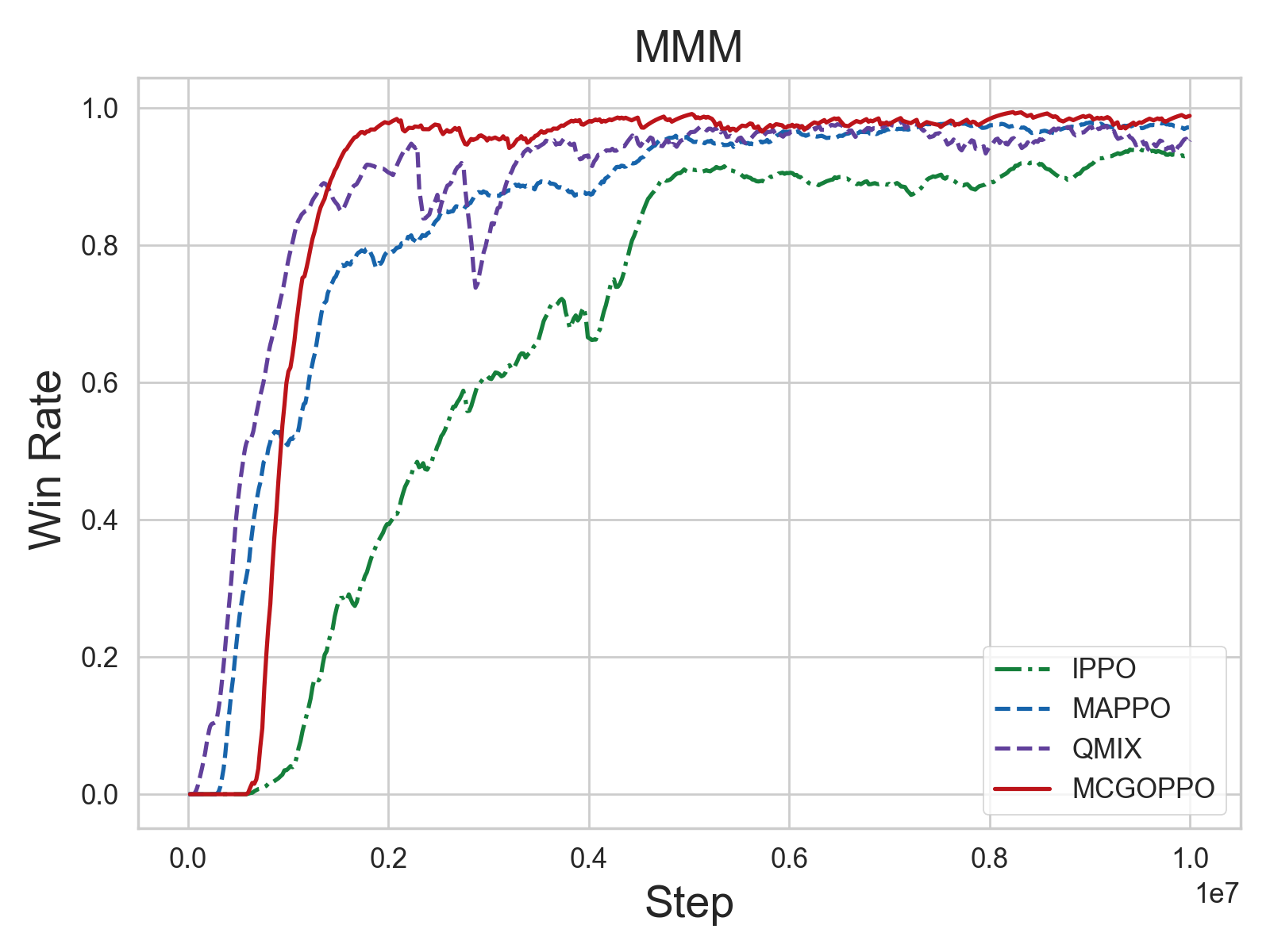}
\caption{Experimental win ratio of scene MMM in SMAC}
\label{MMM_win_rate}
\end{figure}

In the first one million steps, QMIX algorithm grew the fastest, followed by MAPPO algorithm, but MCGOPPO algorithm improved after about one million steps, and MCGOPPO algorithm kept the lead. It can also be seen from the figure that MCGOPPO algorithm converges at about two million steps, while MAPPO algorithm and IPPO algorithm converge at about five million steps. Therefore, it can be considered that MCGOPPO algorithm is superior to other comparison algorithms on the map.

\begin{figure}[h]
\includegraphics[width=10cm]{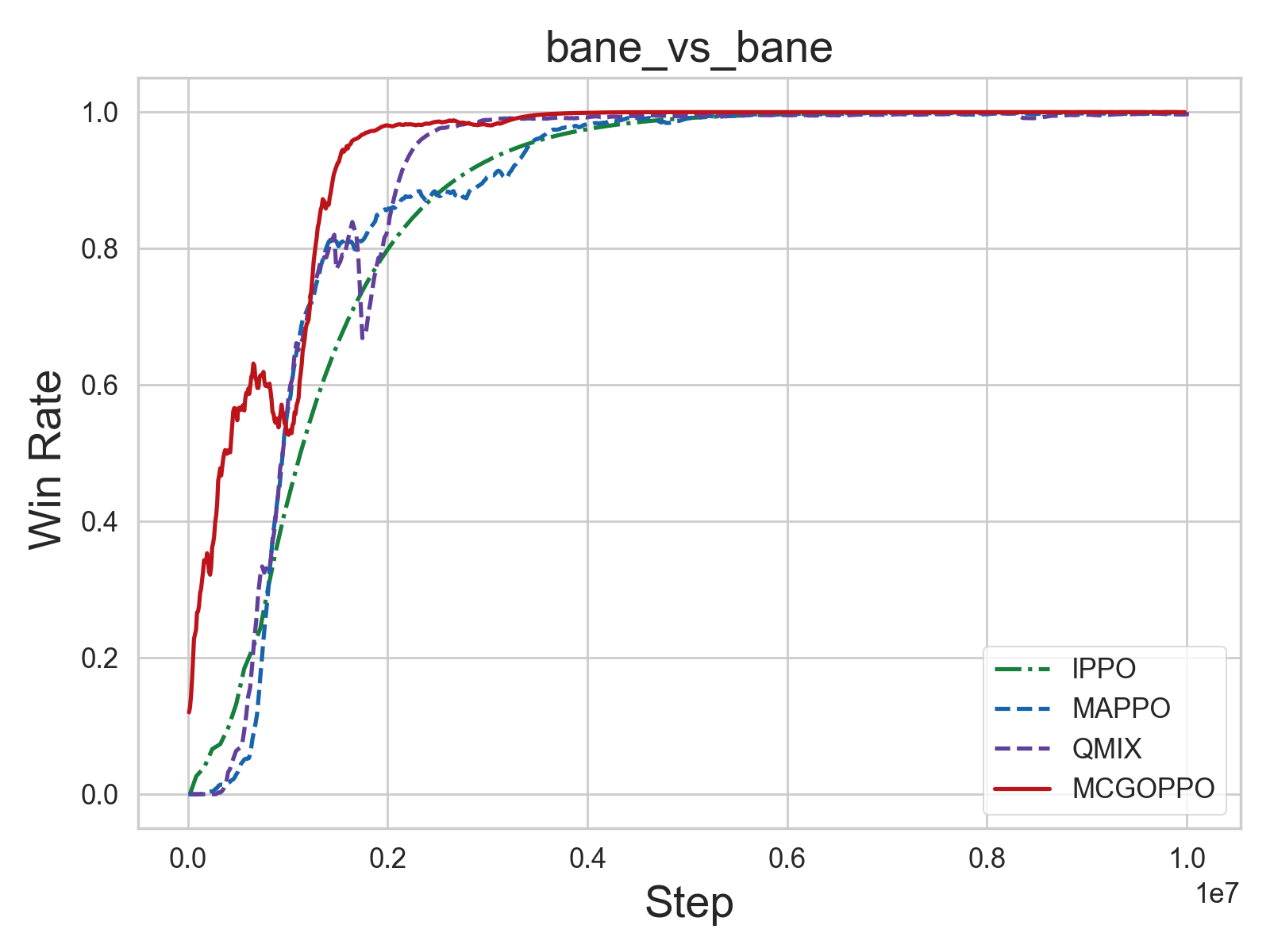}
\caption{Experimental win ratio of scene bane\_vs\_bane in SMAC}
\label{bane_vs_bane_winrate}
\end{figure}

The improved MCGOPPO algorithm has the fastest increase in win rate in the first 700,000 moves, and can reach To about 0.6, but then there is a certain decline, at one million steps back to about 0.54, and then quickly and steadily rise, at two million steps on the basic convergence. At the same time, the winning rate of MAPPO algorithm increases the slowest in the first 700,000 steps, only reaching about 0.1, and then rises rapidly and begins to converge at about 4 million steps. QMIX The convergence of the algorithm is similar to that of MAPPO, but it begins to converge rapidly to 1 at about two million steps. The growth of IPPO algorithm is relatively smooth, which is similar to the curve growth trend of MAPPO algorithm, and also converging at about 4 million steps. Therefore, it can be considered that MCGOPPO algorithm is superior to other comparison algorithms on this map.

In addition, the scene experiment of asymmetric camp maps in SMAC environment is carried out, which are 2s\_vs\_1sc, corridor and 5m\_vs\_6m maps respectively. The map difficulty is gradually increased. Compared with the scene experiment of symmetrical camp maps, the scene experiment of asymmetric camp maps is often more challenging for agents. The experimental results are shown below.

\begin{figure}[h]
\includegraphics[width=10cm]{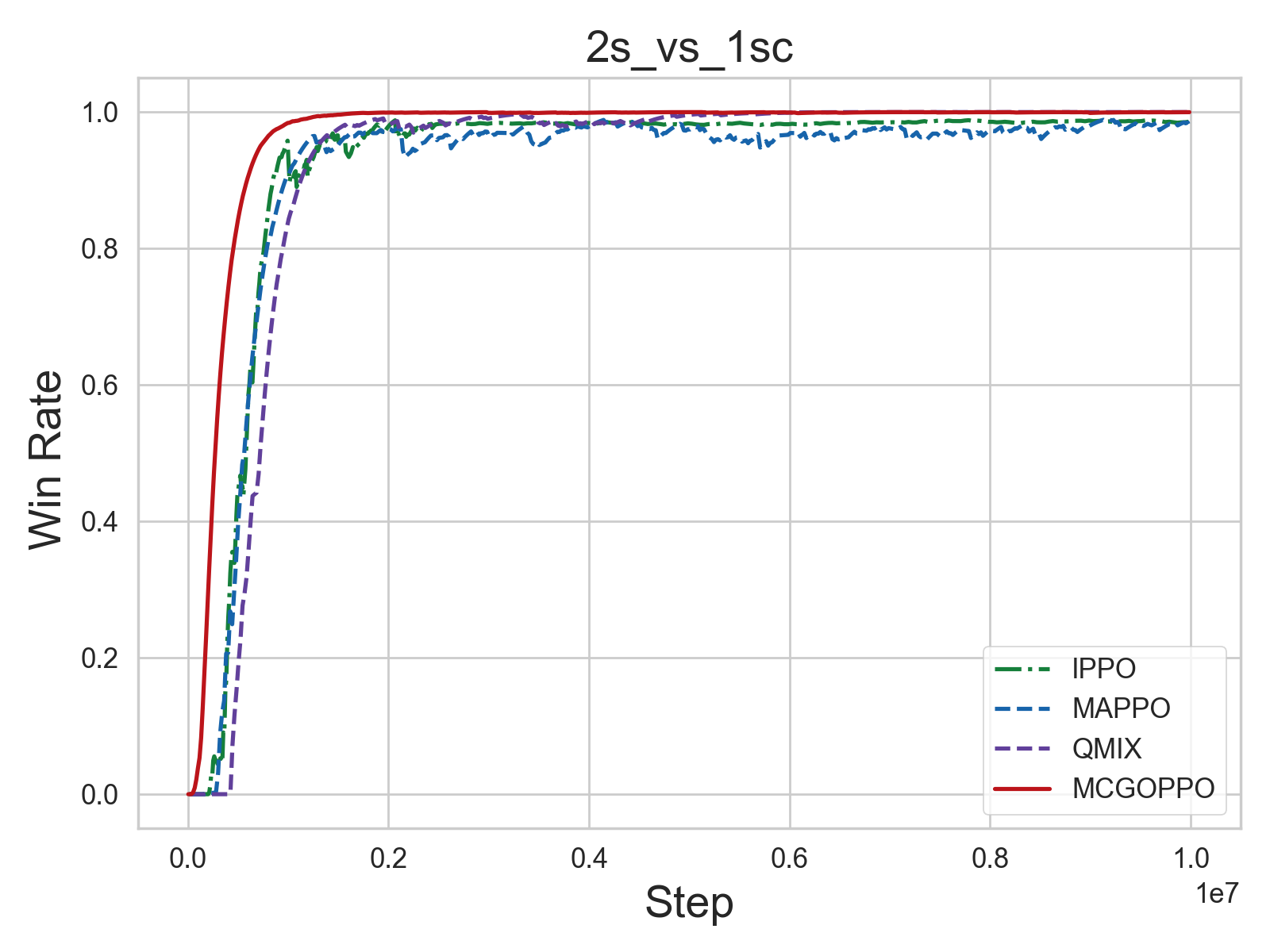}
\caption{Experimental win ratio of scene 2s\_vs\_1sc in SMAC}
\label{2s_vs_1sc_winrate}
\end{figure}

The growth curves of the four algorithms are very similar, and MCGOPPO algorithm converging in about one million steps, while MAPPO algorithm, IPPO algorithm and QMIX algorithm converging in about two million steps, so it can be considered that MCGOPPO algorithm is better than other comparison algorithms on the map.

\begin{figure}[h]
\includegraphics[width=10cm]{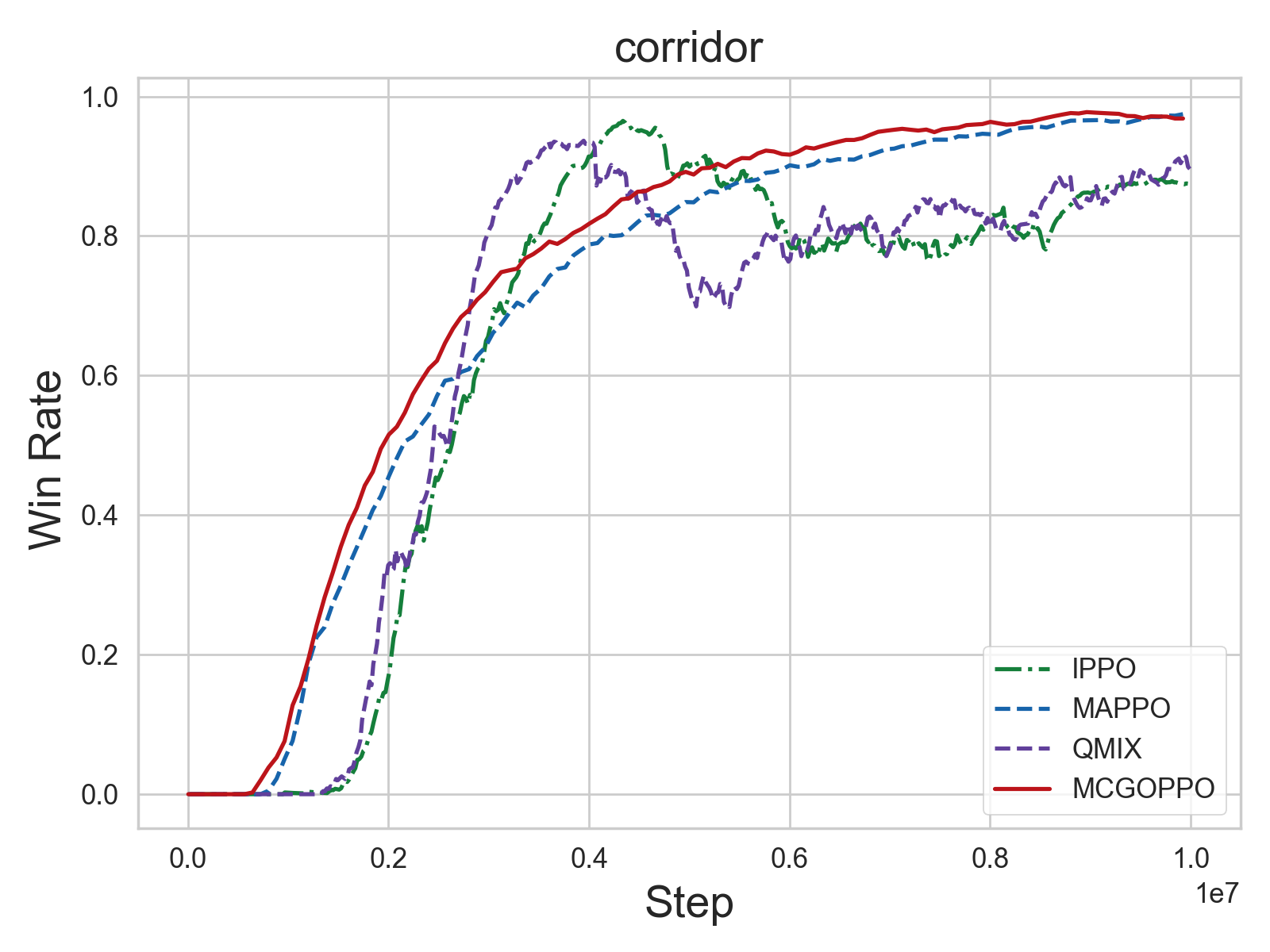}
\caption{Experimental win ratio of scene corridor in SMAC}
\label{corridor_winrate}
\end{figure}

The MCGOPPO algorithm and the MAPPO algorithm have much closer curves, around About 8 million steps basically converges, while IPPO algorithm rapidly rises to around 0.95 in the early stage of about 4 million steps, and then begins to fall back. The convergence curve of QMIX algorithm is similar to that of IPPO algorithm, and finally converges to 0.9. Therefore, it can be considered that MCGOPPO algorithm is similar to MAPPO algorithm on this map, but superior to QMIX algorithm and IPPO algorithm.

\begin{figure}[h]
\includegraphics[width=10cm]{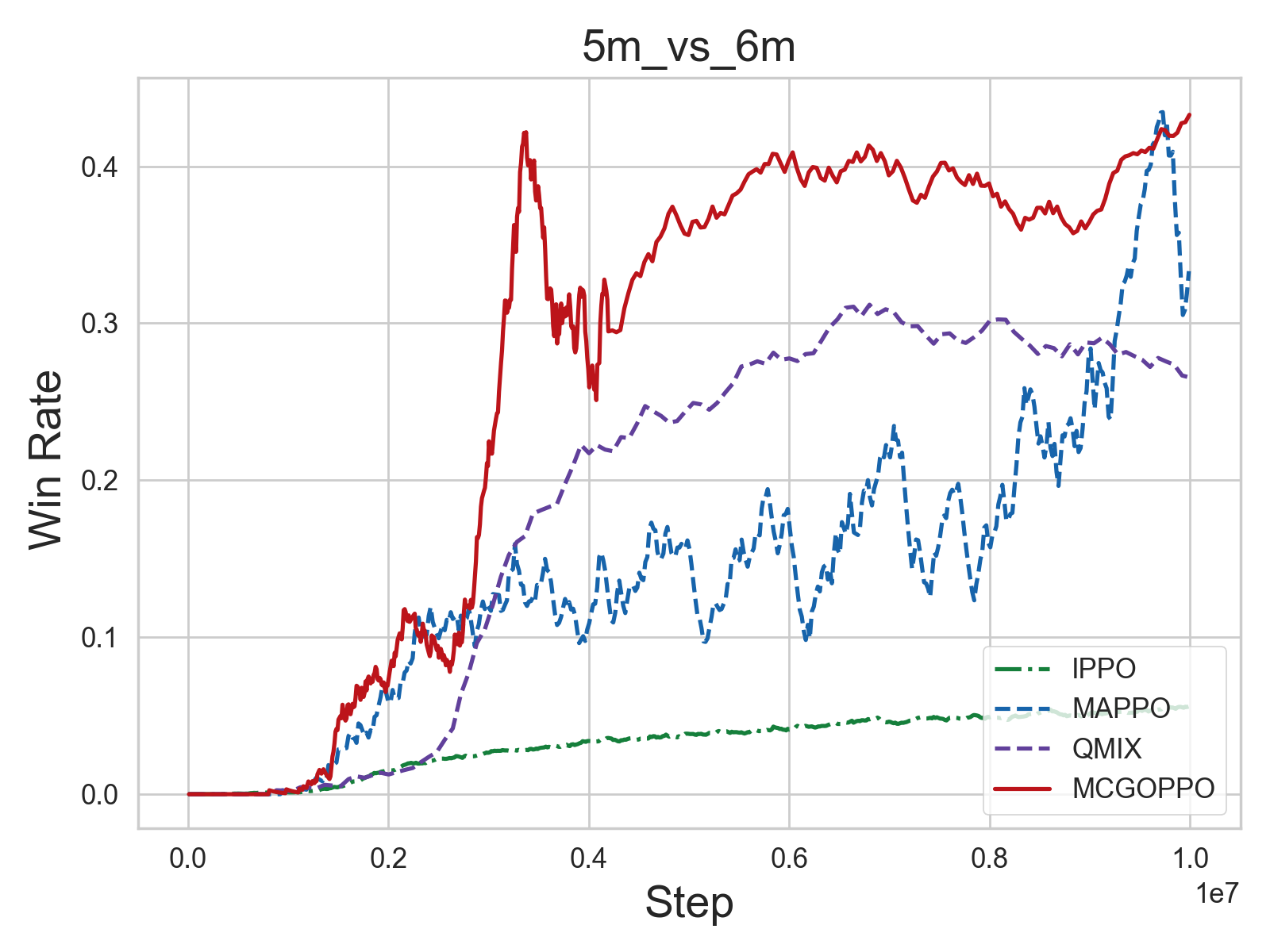}
\caption{Experimental win ratio of scene 5m\_vs\_6m in SMAC}
\label{5m_vs_6m_winrate}
\end{figure}

The MCGOPPO algorithm improves to around 0.4 in about three million steps and then starts to return Drop, at 4 million moves to around 0.25, then slowly rise until the final maximum win rate is 0.44. The growth curve of QMIX algorithm is relatively slow in the first two million steps, and rapidly increases between two million and four million steps. The winning rate reaches 0.32 in seven million steps, then falls back, and finally converges to around 0.27. The growth curve of MAPPO algorithm has been rising in the shock, and finally, in 10 million steps, it is about 0.34, while the growth of IPPO algorithm has been very slow, and in 10 million steps, it is about 0.08, so it can be considered that MCGOPPO algorithm is better than other comparison algorithms on this map.

\section{Conclusion}

Different from the stability of a single agent environment, a multi-agent environment is due to the constant changes of each agent
It has the problem of non-stationarity and brings the challenge to the cooperative decision among multiple agents. The research of this paper is based on the multi-agent environment and adopts the method of deep reinforcement learning to carry out related research. More specifically, based on MAPPO algorithm of CTDE framework, aiming at the problem of lack of information communication between Actor networks in MAPPO algorithm and the problem of redundancy of global information in the input of Critic network, this paper conducted a careful study and proposed an improvement method:

Firstly, the multi-agent communication mechanism based on weight scheduling and attention mechanism is introduced to solve the Actor network
For the problem of lack of information exchange between networks, the non-stationarity of multi-agent environment can be alleviated through information exchange and sharing of agents. The specific approach is to add the communication mechanism to the Actor network, in which the implementation of the communication mechanism consists of two modules, one is the communication selection module of message encoder, weight generator and weight scheduler, and the other is the message processing module composed of attention module. The whole process is: on the one hand, the local observation information of each agent is input into the weight generator to generate the weight coefficient corresponding to the agent, and then the weight is input into the weight scheduler for normalization processing to obtain the final scheduling weight, which is used as the selection basis for the communication between agents later. On the other hand, the local observation information of each agent is input to the message encoder to generate the communication information after compression encoding, which is stored in the message pool as the message for subsequent communication. After the above parallel operation, from the perspective of each agent, the communication information in the corresponding message pool is selected according to the weight in the weight scheduler, and it is taken out and input into the attention module with its own local observation information. Through the processing of the attention mechanism, the information communication between multiple agents is realized.

Second, it introduces global information optimization processing based on attention mechanism and deep and shallow feature processing, because
CTDE framework is characterized by introducing global information to mitigate the impact of non-stationary multi-agent environment during centralized training. However, MAPPO algorithm has redundancy in processing this global information to a certain extent. In MAPPO algorithm, a feature clipping redundancy processing method is further mentioned. However, its operation is artificially preset, which requires the artificial introduction of prior knowledge for feature processing. The improvement method in this paper is to input the joint observation information and global information of all agents into the attention mechanism, and then to obtain the simplified feature information through redundancy processing, and then to carry out the deep and shallow feature processing. For the enemy agent information that is closely related to the target selection, the enemy agent information and the friend information are processed shallow. After that, the features will be spliced and then input into the subsequent centralized Critic network.

\bibliographystyle{unsrt}  
\bibliography{references}

\end{document}